\documentclass[runningheads]{llncs}
\usepackage{graphicx}
\newcommand{\colo}[1]{}

\usepackage{enumitem} 
\usepackage[dvipsnames]{xcolor}  
\definecolor{OliveGreen}{rgb}{0,0.6,0} 
\usepackage[normalem]{ulem}   

\usepackage{fancyhdr}                           
\fancypagestyle{firstpage} 
{
   \fancyhf{}
   \fancyfoot[C]{\vspace{-8mm}\thepage}

   \fancyfoot[L]{\vspace{1mm}\noindent\rule{4.8in}{0.4pt}
     \scriptsize{
         \textcopyright \hspace{0.5mm} Christopher D. Clack and Ciar\'{a}n McGonagle 2018-2019 \\
       This work is licensed under a {Creative Commons Attribution 4.0 International License (CC  BY):}  
       \url{https://creativecommons.org/licenses/by/4.0/} \\ 
       Provided you adhere to the CC BY license, including as to attribution, you are
       free to copy and redistribute this work in any medium or format and remix, transform,
       and build upon the work for any purpose, even commercially.                  \\
       \vspace{1mm}
     }
 }
}
\begin{document} 
\title{Smart Derivatives Contracts: the ISDA Master Agreement and the automation of payments and deliveries}
%
%
\author{Christopher D. Clack
\inst{1}
\and
Ciar\'{a}n McGonagle 
\inst{2}
}
\authorrunning{C. Clack and C. McGonagle}
%
\institute{Centre for Blockchain Technologies, \break 
Department of Computer Science, UCL,\break 
London WC1E 6BT \break
\email{clack@cs.ucl.ac.uk}\\
\and
International Swaps and Derivatives Association, Inc. \break
One Bishops Square, London E1 6AD \break
\email{CMcgonagle@isda.org}\\
}
\maketitle              
\begin{abstract}

%

~~High-value derivatives contracts require substantial legal protection and often utilise standardised legal documentation provided by the International Swaps and Derivatives Association (ISDA).  Smart Derivatives Contracts aim to automate many aspects of high-value contracts, including automation of the provisions of the ISDA legal documentation.  Here we investigate how the ISDA Master Agreement may affect the automation of payments and deliveries: we provide a framework for understanding how high-value derivatives contracts are structured at different levels, in terms of both the legal documentation and the workflow; we explain issues relating to how the smart contract code processes payments-related and deliveries-related events; and we discuss the extent to which 
these
are amenable to automation.

\end{abstract}

\thispagestyle{firstpage} 

\section{Introduction}

A high-value derivatives transaction establishes a financial relationship between counterparties that may last
many decades and may involve a very substantial notional sum.  This requires extensive legal protection and in practice many derivatives transactions utilise legal agreements that are based on standardised legal documentation provided by the International Swaps and Derivatives Association (ISDA).

Smart Derivatives Contracts aim to automate 
these high-value derivatives contracts, including automation of 
lifecycle events stated in the economic terms of the specific derivatives product (such as an Interest Rate Swap)
and the potential automation of aspects of the ISDA Master Agreement.
Such automation is often assumed to be implemented on a distributed ledger, possibly using a blockchain, though many benefits will also accrue from automation on a centralised platform. 
This vision raises many issues to be solved, such as (i) how the smart contract code \cite{stark2016}  (that automates performance) can be faithful to the legal agreement, and (ii) to what extent the provisions of the legal agreement can be automated.  This requires an inter-disciplinary approach that brings together computer scientists, lawyers and banking practitioners.

As we shall explain, much of the operational detail of payments and deliveries can be found in the transaction confirmation and product definitions (i.e. the economic terms and payment mechanics of the particular derivatives product, such as an interest rate swap or an equity swap).  However it is not sufficient just to automate the operational aspects of the contract found in the economic terms; the broader contractual relationship must also be taken into account.  As explained in \cite{ISDA2019}:
\begin{quote}
{\em ``Focusing exclusively on the economic terms of an individual transaction may ignore much of the
external complexity that can affect a party's ability to perform its obligations (or assert its rights) in
relation to that transaction.''}
\end{quote}

\noindent
Here we focus on how the provisions of the ISDA Master Agreement (hereafter ``Master Agreement'') specifically 
affect the automation of payments and deliveries.
We provide a framework for understanding how 
derivatives contracts are structured at different levels,
to assist in understanding the requirements, opportunities and challenges for automation; we explain issues relating to how the smart contract code will need to process payments-related and deliveries-related events; and we discuss the extent to which different kinds of events and processing are amenable to automation.

 \vspace{12pt}
 
Although these discussions are primarily targetted at computer scientists who write smart contract code, we aim to use reasonably non-technical language so that the issues are accessible to specialists from different domains such as lawyers, banking practitioners, regulators, and policy makers. We hope that the issues and views raised in this paper will stimulate debate and we look forward to receiving feedback.

\vspace{12pt}
\noindent 
\textbf{Acknowledgements:} 
We would like to thank Scott Farrell (King \& Wood Mallesons)
for his helpful feedback on an early draft of this paper.

\section{Smart Derivatives Contracts}
\label{sec:smartderivativescontracts}

Smart Derivatives Contracts are smart contracts for automating derivatives contracts.  Here we use the term ``smart contract'' in the sense defined by \cite{SCT2016} and used in \cite{corda2016,SCT-R3-3,clack2018grandchallenge,ClackJDB,SCT2016a,CliffordChanceEBRD2018,harley2017,ISDAKingWoodMallesons2018,ISDALinklaters2017}:

\begin{quote}
{\em A smart contract is an automatable and
enforceable agreement. Automatable by computer,
although some parts may require human
input and control. Enforceable either by legal
enforcement of rights and obligations or via
tamper-proof execution of computer code.}
\end{quote}

This definition captures both the breadth of Nick Szabo's original vision for smart contracts \cite{szabo1997} and the 
breadth of work currently underway across different disciplines.
Although smart contracts are commonly associated with peer-to-peer distributed ledger and blockchain technology, they may also be implemented on a centralised technology platform.  This is an important consideration for the financial services sector, since it is unlikely that all institutions will adopt peer-to-peer technology at the same time \cite{ISDA-CDM-Clack,fintechfutures2018}.

The term ``smart contract'' also has a low-level technical meaning within some distributed ledger technology platforms, where it is used to describe replicated code that runs synchronously on multiple nodes of the distributed ledger.  Where necessary for disambiguation, we use the term ``technical smart contract'' to refer to such low-level code.

Automating the performance of derivatives contracts may allow for a substantial reduction in costs for large financial institutions, for example in terms of streamlined reconciliations \cite{Deloitte2016,ECB2016,fintechfutures2018,ISDA2016}, reduced numbers of intermediaries \cite{Deloitte2016}, reduced staffing requirements and reduced human error \cite{BIS2017,ECB2016}.  Other aspects of automation occur in terms of streamlining the pre-trade workflow and in the processing of disputes, using a blockchain to provide an immutable history of versions of legal documentation.  Through the use of Smart Contract Templates \cite{SCT2016,SCT2016a} the testing and debugging of code can be achieved much earlier in the code lifecycle,\footnote{Which reduces cost, and accelerates downstream processes.} and through the use of semantic analysis of the legal documentation \cite{SCT-R3-3,ClackJDB,ClackVanca2018} test cases can be constructed and used for code validation.

In the remainder of this section we discuss relevant aspects of the ISDA document architecture that provides {\em standardisation} (a necessary precursor for automation), and the {\em smart contract code} that directs the automated performance of a Smart Derivatives Contract.

\subsection{Standardisation of Smart Derivatives Contracts}

OTC derivatives are often purchased 
as a mechanism for risk management so that the precise form of the purchased derivatives product will match the purchaser's financial exposures.  
Derivatives play an important role in helping to reduce the uncertainty that comes from changing interest rates, and exchange rates, as well as 
credit, commodity and equity prices.  Derivatives are used by thousands of companies in all industries and in all regions \cite{ISDA-who-uses-2017}.
These derivatives contracts can have substantial value, complexity and longevity,
and a firm that purchases a bespoke derivatives product will need legal clarity and protection relating to the terms of the agreement.  

Negotiating the terms and conditions of bespoke derivatives contracts can itself be a lengthy and costly process, and this complexity and cost can be improved through standardisation.  
Over the past 30 years, ISDA has worked with its members to produce and maintain 
a documentation framework to support the growth of the derivatives industry. This has been achieved through the development of the ISDA Master Agreement and 
corresponding documentation, such as various annexes,\footnote{For example, a credit support annex may be added setting out the terms on which collateral is exchanged between the parties to reduce credit risk.} definitional 
booklets\footnote{For example the 2006 ISDA Definitions, which provide the basic framework for the documentation of interest rate and currency derivatives tranactions.} and protocols.\footnote{For example the 2013 EMIR Portfolio Reconciliation, Dipute Resolution, and Disclosure Protocol, which enables parties to amend their agreements to reflect certain requirements of the European Market Infrastructure Regulation.}
 This framework has provided legal certainty, clarity and efficiency for 
derivatives market participants.  

The Master Agreement \cite{ISDAUserGuideMA} is central to the ISDA documentation architecture; it 
can govern multiple 
derivatives transactions, the economic terms of which are documented in separate Confirmation documents (written or electronic), which each form part of the relevant Master Agreement. 

The Master Agreement has been structured as a complete contract, containing payment provisions, representations, agreements, 
events of default, termination 
events, early termination provisions, methods for calculating payments on early termination and other provisions. 
These terms apply to each of the derivatives transactions 
entered into under that Master Agreement, such that all transactions are governed by a single agreement comprising the set of related documents.

The Master Agreement is produced in a standard, pre-printed form and is not intended to be directly amended by the parties. 
It is however possible to amend the terms of the pre-printed form by completing the Schedule to the Master Agreement. 
In the Schedule, parties may choose whether and how certain operative provisions within 
the Master Agreement will apply.\footnote{For example, in Part 1(a) of the Schedule, parties may choose to define
``Specified Entities'' to expand the scope of certain Events of Default and one of the Termination Events.}  
Also in the Schedule, parties may alter or amend the provisions of the Master Agreement as 
they wish through specification of additional or alternative provisions.

Any changes made to the Schedule will, unless stated otherwise, apply to all derivatives transactions entered into under the relevant Master Agreement. Further changes may also be made in the Confirmation for an individual Transaction; such changes will only apply to that Transaction.

Depending on counterparty type and product complexity, it may be the case that certain Master Agreements are subject to a high 
degree of customization. As explained in \cite{ISDA2019}:\footnote{For further discussion of the potential complexity of derivates transactions under the ISDA document architecture, the reader is referred to \cite{ISDA2019}.}

\begin{quote}{\em
If a contractual term has been included in a transaction
confirmation that conflicts or is inconsistent with a term in the Master Agreement (or Schedule),
then the relevant term in the transaction confirmation will take precedence
}\end{quote}

\begin{figure}[h]
\centering
\includegraphics[width=13cm, trim={1cm 4cm 2cm 2cm}, clip]{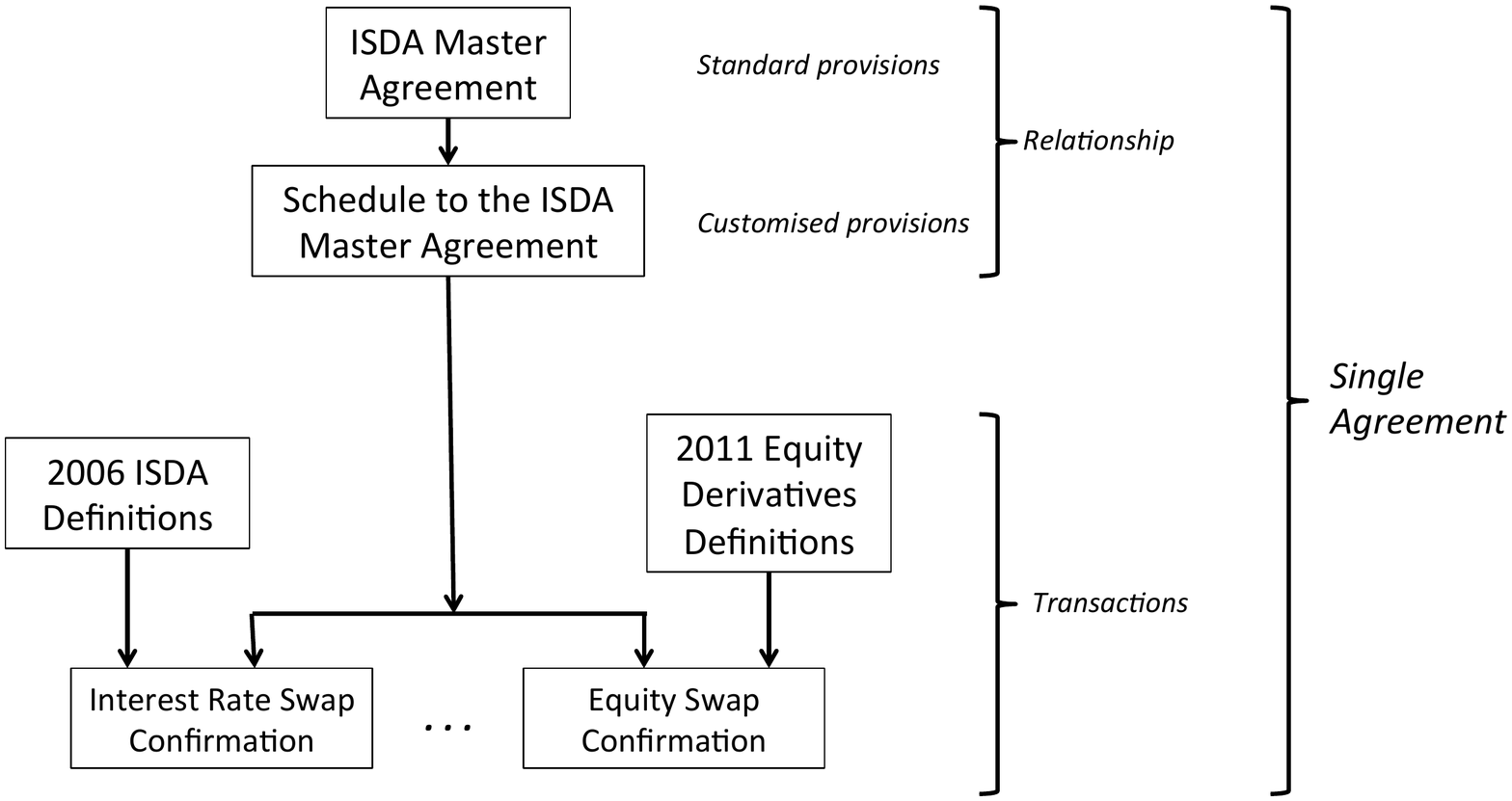}
\parbox{10cm}{\caption{\label{fig:ISDADocArch} The ISDA Document Architecture.}}
\end{figure}

\noindent
Figure~\ref{fig:ISDADocArch} illustrates the ISDA document architecture where a single Master Agreement is coupled with a negotiated Schedule to create
a legal relationship between the parties that governs a potentially large number of Transactions, each of which may be a different type of derivatives product.  The figure illustrates two such Transactions, and shows how the Confirmation document for each draws upon the relevant ISDA Definitions to
assist in defining the economic terms and mechanics for each product.  All of the Transactions are part of a single agreement --- this is important 
because it facilitates
issues such as payments netting and close-out netting.

There are a number of different versions of the Master Agreement, including the 1992\footnote{There are two versions of the 1992 ISDA Master Agreement: the ``local 
currency -- single jurisdiction'' version and the ``multicurrency -- cross border'' version.} and 2002 versions. This paper will focus primarily on the 2002 version of the Master 
Agreement and throughout this paper the word ``Section'' refers to a Section of the 2002 Master Agreement unless otherwise specified. 
However, many of the concepts and issues discussed will be common across each of the different versions.

\subsection{Smart Contract Code}
\label{sec:smartcontractcode}

Smart Derivatives Contracts aim to automate derivatives transactions, and an important aspect of this is to automate the performance of 
those transactions by capturing the logic of the agreement in computer code  --- ``smart contract code'' \cite{stark2016}.  
This includes the capture of deontic logic (rights, prohibitions and obligations), operational logic (actions), business logic 
(e.g. calculations), and temporal aspects (expressions relating to time).
 
To a greater or lesser extent (as discussed later), the smart contract code should address the following:
\begin{itemize}
\item
The logic and data contained in the Master Agreement, the negotiated Schedule, and possibly other documents from the ISDA documentation architecture.
\item
The logic and data contained in the Confirmation document that is the subject of each Transaction.
\item
The logic and data contained in the ISDA definitions for the derivatives instrument that is the subject of each  Transaction.
\end{itemize}

\noindent
In a joint white paper released in 2017 \cite{ISDALinklaters2017}, ISDA and Linklaters distinguished between two different models of performance automation: the ``external model'' and the ``internal model'':
\begin{description}
\item[The external model] logically separates the smart contract code from the smart legal contract; the code is not part of the legal agreement and is merely one way in which the parties may perform their rights and obligations under the agreement.  Thus the code has no legal effect.  Although many aspects
of contract performance are currently (separately) automated,  {\em smart contract code} would add the novel aspect of {\bf autonomous} operation, where the code takes higher-level control over the activities performed, makes decisions based on observed events, and has far less need for human intervention.\footnote{We note in passing that it is not true that only provisions that are ``formally specified'' can be automated --- simple
aspects of derivatives contracts are already performed automatically (though perhaps not autonomously), and it is trivially true that simple
provisions written in English can be performed automatically by computer.}
\item[The internal model] envisages that all or some of the autonomous smart contract code (written in some formal representation) would be part of the contract and would have legal effect:\footnote{An extreme version of the internal model might be for all of the contract to be represented formally, though many problems would have to be solved first (such as how to embody in a formal representation the degree of ambiguity and flexibility that is currently used in
legal text).  Current research in this area --- including unpublished work currently underway at UCL --- often uses the term ``computable contracts''.}
\begin{quote}{\em
%
Certain clauses would
be drafted in natural human language, as is the case today. But other clauses would effectively be
set down on the page in some form of code, or other formal representation. Alternatively, instead
of setting down the code or formal representation within the written contract itself, the written
contract could refer to an identified piece of code stored elsewhere and could state that such code is
to be given legal effect between the parties. \cite{ISDALinklaters2017}
}\end{quote}
\end{description}

\noindent
The final observation in the above quote is important --- that the code need not be constrained to snippets of
``formal representation'' dispersed through the contract, but may be stored elsewhere.  Although some kinds
of formal representation might be suitable for presentation as separate small pieces,\footnote{For example,
if represented using a mathematical or logic notation, or using a declarative programming language
such as Haskell \cite{jones2003haskell} or Miranda \cite{clack1995programming,turner1985miranda}.}
many are not; for example,
with imperative programming languages (those consisting of a sequence of commands), the order in which
commands appear can strongly influence the meaning of the code and it is not always possible to impute
an unambiguous meaning to a small piece of such code.\footnote{This is similar to the difficulty observed in
attempting to understand the meaning of a single legal provision without reading the entire contract.}  For this
reason, and others, it is possible (perhaps even likely) that the ``formal representation'' used in the internal 
model would not be the same as that used for the final smart contract code that automates performance.\footnote{A
related observation is made in \cite{ISDAKingWoodMallesons2018}, that translation might occur in two steps:
first from English text to a formal representation at an intermediate level (using a Domain Specific Language), 
and then in a second step from the
formal representation to smart contract code.  A Domain Specific Language would provide both a formal
definition of the legal contract and a formal definition to control the production of the smart contract code.}

The internal model has the potential to bring significant additional benefits to 
Smart Derivatives Contracts, for example in improving the fidelity of the smart 
contract code to the contract.  
However, both the external and internal models will be equally subject to the
issues discussed in this paper, and so we stick to the simpler of the two --- hereafter we assume the external model.


\subsubsection{Smart Contract Templates.}
\label{sec:smartcontracttemplates}

UCL and Barclays have developed ``Smart Contract Templates'' \cite{SCT-R3-3,ClackJDB,SCT2016,SCT2016a} to encourage code development --- including verification (testing, debugging) and validation --- to be aligned with the workflow associated with the ISDA documentation architecture, and to occur as early as possible in the lifecycle of the code.  The name ``template'' indicates that the code will leave some terms as yet undefined.\footnote{Smart Contract Templates should not be confused with the very
low-level function ``templates'' mentioned in \cite{ISDAKingWoodMallesons2018} (which are discussed further in Section~\ref{sec:howmuchtoautomate} of this paper).}  These templates may be developed and comprehensively verified and validated in advance (using example values for as yet undefined terms), and this might include different versions of the code designed to run on different technology platforms.  
The envisaged steps are illustrated in Figure~\ref{fig:SCT} and further explained below:

\begin{figure}[p]
\centering
\includegraphics[width=12.2cm, trim={2.2cm 0.8cm 4.1cm 1cm}, clip]{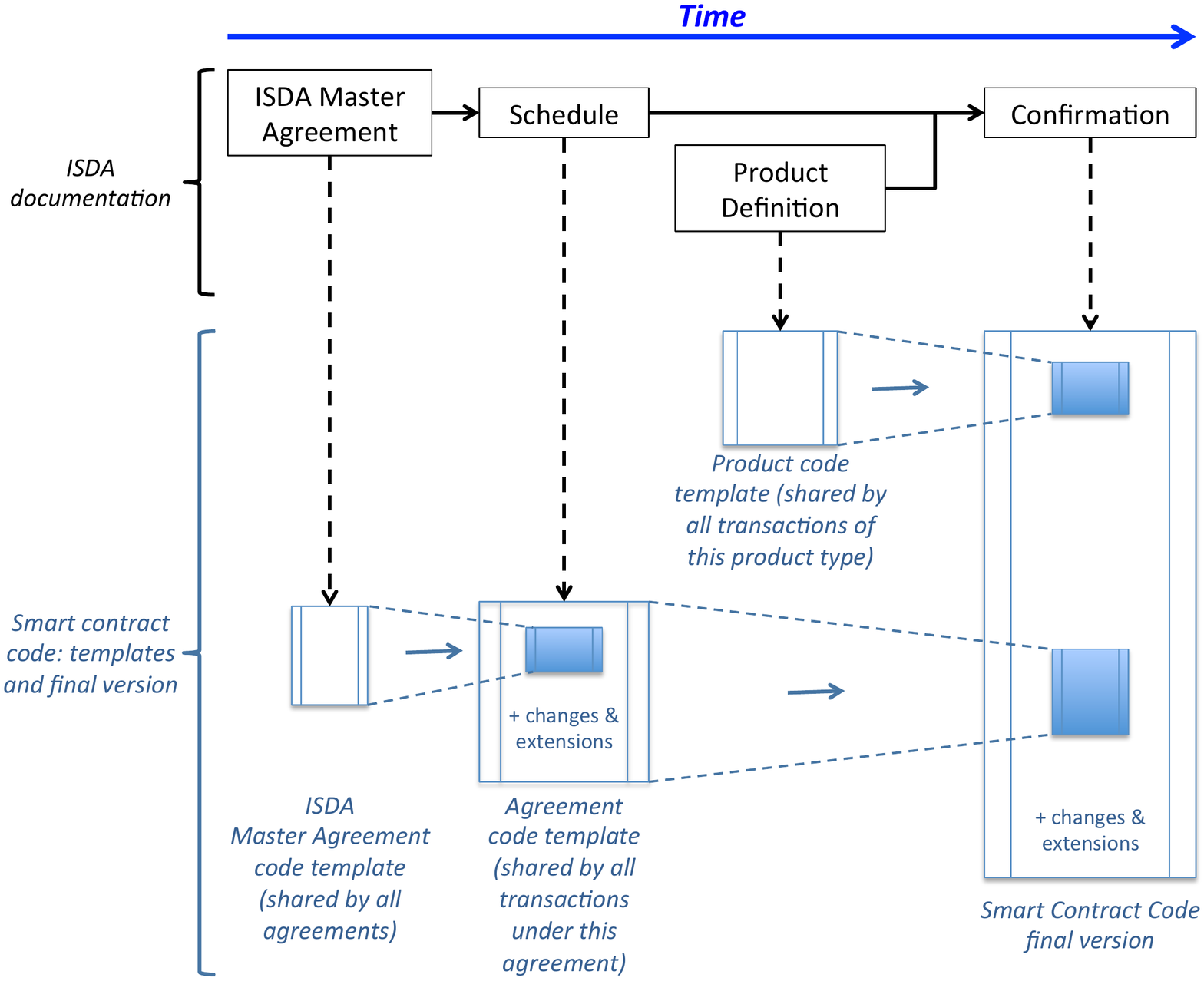}
\parbox{12.2cm}{\caption{\label{fig:SCT} Smart Contract Templates align the workflow of construction, verification and validation of smart contract code with that of the ISDA document architecture, moving verification and validation to as early as possible in the software lifecycle and maximising sharing.  The top of the diagram illustrates the progression through time of use (from left to right) of the ISDA documentation from Master Agreement to negotiated Schedule, via Product Definition to a Confirmation for a Transaction.  Below, there is a matching progression through time of use of the Smart Contract Templates, each separately verified and validated before being incorporated into the next,  and producing the final version of the smart contract code.  The vertical dashed arrows indicate how each ISDA document is used to derive its corresponding code.}}
\end{figure}

\begin{enumerate}
\item
Prior to any negotiated agreement between parties, smart contract code should be developed in ``template'' form to capture the logic and data of the Master Agreement, and a wide range of defined derivatives products.  This creates (a) a Master Agreement code template, shared by all agreements that are based on this Master Agreement, and (b) a number of Product Definition code templates, one for each derivatives product type and shared by all Transactions of the same type.

\item
When two or more counterparties establish an agreement, based on the Master Agreement, they will often negotiate modifications that are captured in the Schedule.  When the negotiation is complete, a copy will be made of the existing Smart Contract Template mentioned in 1(a) above and the code in this copy will be modified --- many of the undefined terms in the template will be given appropriate values, and depending on the extent of the modifications to the legal provisions this may also require a more or less substantial rewriting of the smart contract code.  
The code is likely to require further verification and validation, and this process will substantially benefit from the fact that the previous template had already been verified and validated.

\vspace{6pt}
The resulting modified code should accurately reflect the intentions of the parties 
under the agreement, but it is not yet ready to run since the parameters
for individual transactions are not yet known. Hence, this is still a code ``template''; we call this the Agreement code template.
\item
For each executed transaction, the template mentioned in point 2 above is copied and extended to include the logic and data contained in or referenced from within the Confirmation document; this will include a variety of transaction parameters that are currently undefined in the template, together with the appropriate Product code template (from point 1(b) above).\footnote{Additional parameters might also be passed to the code, e.g. a unique identifier that can be used to retrieve the original signed legal documents in the case of dispute.}  
The final version of the
smart contract code is then ready to be instantiated and run on a centralised or distributed ledger technology platform.
\end{enumerate}

\noindent
To ensure financial services sector confidence in the accuracy and reliability of smart derivatives contracts, it is essential that the final version of the
smart contract code 
behaves in a manner that is faithful to the legal agreement.  This is not straightforward, since the semantics of  legal agreements and economic terms can be complex, since the disciplines of law and computer science have conflicting definitions of seemingly innocuous terms \cite{ClackJDB},
and since a na\"{i}ve technical implementation of some processes may not be legally effective \cite{ISDAKingWoodMallesons2018} or may cause 
unexpected or undesirable legal or contractual issues.

The validation of the smart contract code (to ensure that the code is faithful to the legal contract) is of paramount importance, and strongly motivates our discussion here of those aspects of the ISDA documentation that must be considered by the technology practitioners who will develop, verify and (together with lawyers) validate the smart contract code.  
This includes discussion of the Master Agreement, especially key aspects of the Master Agreement that directly affect the operation of payments
and deliveries.

\newpage
\section{A framework for understanding events in Smart Derivatives Contracts}

ISDA documentation establishes rights
and obligations and agreed procedures that can affect both the timing and quantum of payments and deliveries
for a potentially very large number of transactions.  For example,
at a {\colo{red}\sout{low}detailed operational} level the economic terms of an individual transaction direct the payments and deliveries that are applicable only to that transaction;
and at a {\colo{red}\sout{high}more strategic} level these lower level payment obligations can be aggregated and netted to generate a smaller number of smaller-quantum
net payment obligations.

The
processing of payments and deliveries throughout the lifetime of a derivatives transaction can be 
affected by different kinds of event, and 
the provisions of the ISDA documentation (especially the Master Agreement) 
guide the responses to such events.  
The event or circumstance needn't relate directly or specifically to the derivatives transactions between the 
parties. Indeed, there are scenarios where an Event may occur where the parties are continuing to meet their 
payment and delivery obligations under all of their derivatives transactions and where there is no immediately 
apparent reason to suggest that they may no longer be capable of doing so.

In order to facilitate 
understanding, and to support the aim to automate 
a broad range of aspects of Smart Derivatives Contracts,
we provide in Section~\ref{sec:events} an overview of events and in Section~\ref{sec:levels} a framework 
based on an analysis of the different levels at which events can occur.

\subsection{Overview of Events and potential Events}
\label{sec:events}

In this paper we make a distinction between an ``event'' (which in our context is any thing or circumstance that is observable by the 
smart contract code that performs the derivatives contract) and an ``Event'' (which is an event specifically defined in the Master Agreement).

In the Master Agreement an ``Event'' is an event or circumstance that may impact upon 
the parties' respective ability to perform their obligations, including payment and delivery obligations, under 
the derivatives transactions entered into between them.  
Thus, the set of possible Events is a subset of the set of possible events.

It is important that parties are able to react to events which may be indicative of either
a deterioration in creditworthiness of their counterparty or some fundamental change
in the legal, regulatory or operating framework in which their counterparty is operating
such that their ability to continue making payments and/or deliveries could be impeded.
The Master Agreement therefore contemplates the occurrence of  broad range of such events
and provides each party with a mechanism to terminate derivatives transactions in order to
eliminate or mitigate its financial exposure to its counterparty.

The Master Agreement contemplates two distinct types of Event:  Events of Default and Termination Events.
Broadly speaking, an Event of Default might arise where one of the parties is considered 
to be at fault.\footnote{For example, a Failure to Pay or Deliver (Section 5(a)(i))}
Termination Events are different in that they are intended to capture events where neither party is strictly at 
fault.\footnote{For example, an Illegality (Section 5(b)(i)) where an action is observed to be illegal 
(even though previously it may have been legal). }

The Master Agreement provides for a standard number of 
Events of Default
and Termination 
Events.
All of these Events are capable of further customisation within the Schedule.
One should therefore not assume  
that a particular scenario would give rise to an Event of 
Default or Termination Event under every Master Agreement in the market (or that the consequences of 
any such Event of Default/Termination Event would be the same across every Master Agreement).

The Master Agreement also
contemplates that the parties may agree ``Additional Termination 
Events'' to capture potential or perceived risks that might arise with respect either to the derivatives transactions 
entered into, or to their counterparty, or to the broader market and regulatory 
framework.\footnote{The Master Agreement does not explicitly contemplate the inclusion of additional 
Events of Default.  However, there is nothing preventing the parties from agreeing to include additional 
Events of Default should they wish to do so.}

While the ultimate consequence of the occurrence of either type of Event is the same i.e. the potential 
termination of derivatives transactions entered into between the parties, they are necessarily distinct. For 
example, while the occurrence of either type of Event gives a party the potential right to terminate derivatives 
transactions entered into under the Master Agreement, the manner in which these derivatives 
transactions terminate may differ depending on whether an Event of Default or Termination Event has 
occurred.\footnote{See for example the differences between Section 6(a) (Right to Terminate Following Event of Default) 
and Section 6(b) (Right to Terminate Following Termination Event).}

Specific situations must be considered.  For example:
\begin{itemize}
\item
Where an event gives rise to more than one type of Event.
In such cases, the Master Agreement provides a hierarchy for determining how the Event should be 
treated.\footnote{Section 6(c).}
\item
Where
the same Termination Event might exist concurrently with respect to both 
parties. For example, an event might occur that results in it becoming simultaneously illegal for both parties 
to continue meeting their respective obligations under the  Master Agreement. In such scenarios, both 
parties would be ``Affected Parties'' for the purposes of determining how the termination 
mechanics under the  Master Agreement would operate.
\end{itemize}

\noindent
Understanding and correctly categorizing the relevant Event is therefore important for understanding the 
contractual consequences that flow from the occurrence of either of these types of Event
and the precise manner in which an Event might impact upon the parties' respective payment and delivery obligations.

Any technology solution that intends to automate payments and deliveries under a Master Agreement will 
need to take account of the types of event that might occur and be capable of:
\begin{enumerate}[label=\alph*)]
\item
Observing the 
occurrence of a circumstance that might give rise to an event;
\item
Determining that an event has occurred; and
\item
Take action to manage the consequences that might arise from the occurrence of the event.
\end{enumerate}

\noindent
The ability to customize also means that there could be scope for modifying some of these events to 
make them more "automation-friendly." For example, by removing subjective elements in a
particular clause, 
harmonising grace periods etc.  This would need to be done carefully so as to avoid disrupting 
other important contractual provisions within the Master Agreement
or any other relevant agreement.


\subsection{Overview of the different``levels" at which payments-related and deliveries-related events can occur}
\label{sec:levels}

While the Master Agreement (and ISDA documentation generally) is largely standardized, certain 
contracts can often be quite heavily negotiated and customized. The ``contract'' relating to derivatives 
transactions between two parties is often represented by a combination of documents.  These documents are 
highly interdependent. It is not possible to fully understand a single derivatives transaction or the overarching 
contractual relationship between the parties simply by looking at an individual transaction Confirmation or 
even by reference to the Master Agreement.

Furthermore, certain Events under the Master Agreement can only be observed by reference to other 
documents to which one or both parties is a party. For example, where a party's obligations under the 
Master Agreement are supported by external credit support or guarantee, a failure in the efficacy of that credit 
support may constitute a Credit Support Default.\footnote{For example, a failure to maintain any security interest granted to the other party, or the unanticipated cessation of a financial guarantee provided by a third party.}
Observing the occurrence of this Event requires that the 
parties observe the terms of other contracts. 

To fully understand the terms of a particular transaction and how external events may impact upon it, it is 
important to look at each of the various levels of obligation that exist within the ISDA documentation 
architecture, the key documents involved, and how they interrelate.

It is possible to distinguish four different categories or levels at which circumstances or events might 
occur and which may ultimately give rise to the occurrence of an ``Event'' that, in the context of the Master Agreement, could be an Event of Default or a Termination Event:
\begin{enumerate}
\item
By reference to the derivatives transactions entered into between the parties (the Transaction Level) 
\item
By reference to the broader contractual relationship between the parties (the Relationship Level)
\item
By reference to one or more third parties (the Third Party Level)
\item
By reference to an external event that is not directly related to specific derivatives transactions 
entered into between the parties nor to their broader contractual relationship (Exterior Level)
\end{enumerate}

\begin{figure}
\centering
\includegraphics[width=12cm, trim={2.4cm 1cm 8cm 1cm}, clip]{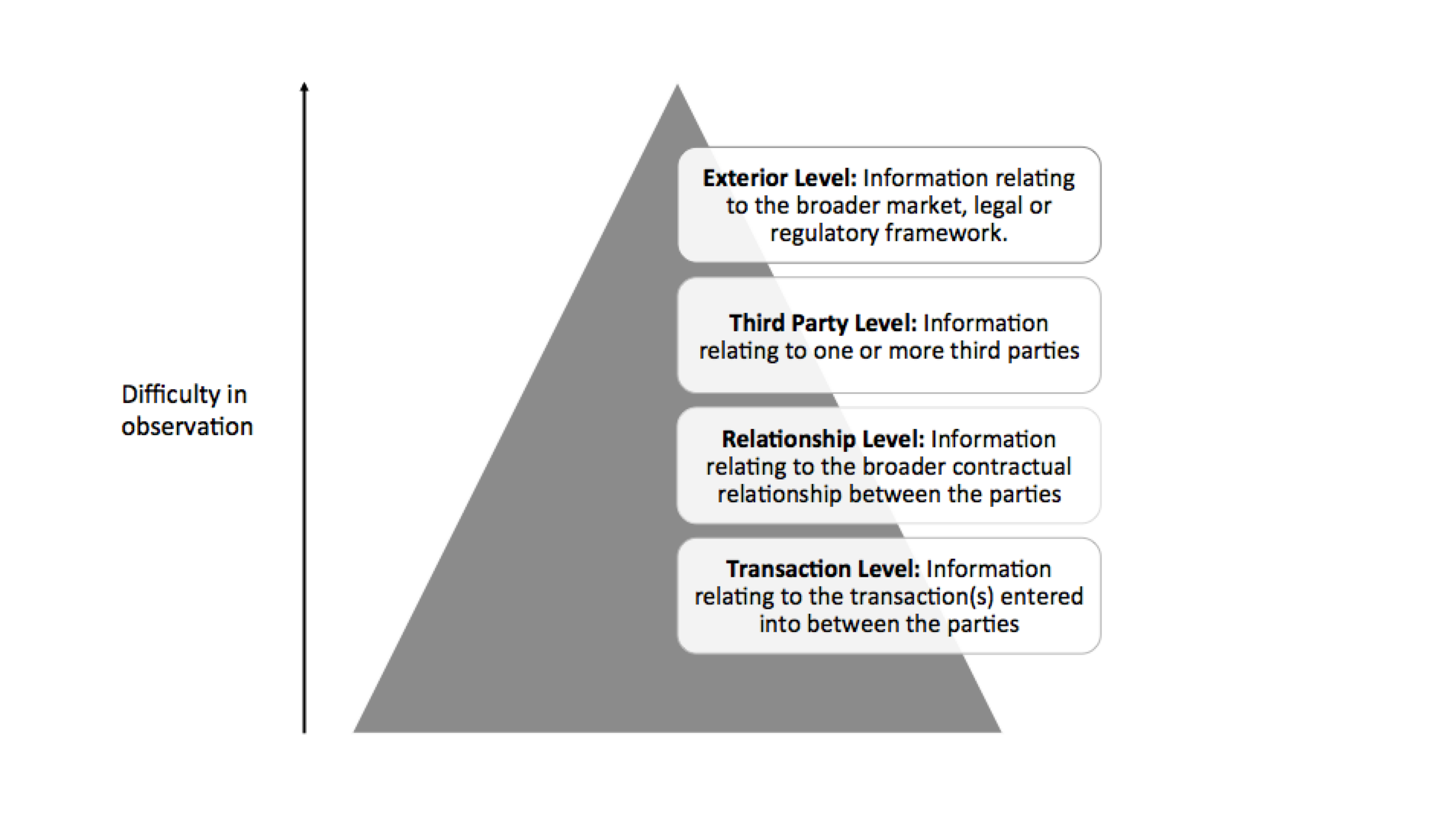}
\parbox{10cm}{\caption{\label{fig:levels} The four levels at which events may occur. }}
\end{figure}

\noindent
Due to the nature and type of information that must be observed to allow the determination of whether an Event has taken place, 
the relative difficulty of observation of events differs between these levels.  Figure~\ref{fig:levels} illustrates the type of information
that would need to be observed at each level and indicates the increasing difficulty in observation between these levels. 

\vspace{-12pt}
\subsubsection{Level 1: Transaction Level.}
Events occurring at the Transaction Level are typically related to the specific product lifecycle, with expected behaviour being set 
out in the Confirmation and product definitions.  Observing
the occurrence of an event at  this level would seem to present the fewest challenges for smart contract code. 
For example, it should
be relatively straightforward for the code to determine whether a party has failed to make a required payment of the required amount at the required time, given that it will have immediate access to the relevant transaction data.  

\subsubsection{Level 2: Relationship Level.}
Events occurring at this level are related to the agreement negotiated between the counterparties and may 
involve more than one transaction (for example, the Master Agreement permits payment netting, which can aggregate lower-level payment obligations from a number of transactions and generate a smaller number of smaller-quantum net payment obligations).

The occurrence of an Event at the Relationship Level may require the smart contract code to look beyond the transaction data and 
to the broader legal and contractual relationship between the counterparties.
For example, when entering into the Master Agreement, the parties exchange a series of 
representations. One of these representations provides that both parties represent to each other that they are:

\begin{quote}
{\em
``duly organized and validly existing under the laws of the jurisdiction of its organization or incorporation and, if 
relevant under such laws, in good standing;''\footnote{Section 3(a)(i).}
}
\end{quote}

\noindent
Whether or not a party is validly existing under the laws of the jurisdictions in which it is incorporated is not 
something that can be ascertained through observation of transaction data. In order to determine the 
accuracy of this representation, it would therefore be necessary for the smart contract code to obtain and review information 
relating to the incorporation of the counterparties, presumably from a public registry.\footnote{Section 4(a) provides that parties must provide to each other any documentation specified in the Schedule to the Master Agreement or in any transaction confirmation. Parties often agree to deliver constitutional documentation as a means of evidencing the validity of their incorporation.  An alternative for contract automation would be to give express permission for the smart contract code to retrieve the relevant documentation from a public registry.}

Events of Default and Termination Events (see Section~\ref{sec:events} of this paper) are examples of Events that might be triggered due to an event which can be observed by looking to information about the parties themselves rather than at transactional details.  For example, the smart contract code might be able to observe Bankruptcy of a party by monitoring information sources that might publish information relating to the insolvency of that party (e.g. a regulatory authority or similar administrative, regulatory or judicial body).

\subsubsection{Level 3: Third Party Level.}
Observing the occurrence of potential Events at the Third Party Level is likely to be more challenging. Here, 
the smart contract code may be unable to establish the potential occurrence of an Event by reference to either the derivatives 
transaction data or to information relating solely to the counterparties. Instead, the code will need to observe 
information relating to a third party i.e. a party who is not a contracting party to the Master Agreement.\footnote{Initially this might be difficult for smart contract code to observe.  However, as the number of automated products and transactions increases, this information might become easier to observe (presumably with permission from the
observed party).}

An example of such an Event is Cross-Default.\footnote{Section 5(a)(vi).} 
A Cross-Default might occur where there has been a default 
by one of the parties under any obligation in respect of Specified Indebtedness.\footnote{``Specified Indebtedness'' is defined in the Master Agreement 
as any obligation in respect of borrowed money.  In practice, this is sometimes amended by the parties so that it extends beyond
obligations in respect of borrowed money (e.g. to other trading exposures) or is narrowed to exclude some obligations (e.g. bank deposits).}
The Cross-Default Event of Default usually also contains a threshold amount, below which defaults in respect of 
Specified Indebtedness would be deemed insufficient to constitute an Event of Default. 

For smart contract code to automate the monitoring of such events, it would be necessary for the code to 
obtain information and observe any and all such arrangements relating to 
Specified Indebtedness entered into between each of the parties and a third party and the precise quantum of 
indebtedness under each arrangement in order to determine whether this Event may have occurred. 
Although the smart contract code is unlikely to have access to {\em all} such arrangements, we could envisage
a future technology ecosystem where large amounts of such information may be observable.

It may even be necessary to look to those arrangements in respect of 
Specified Indebtedness entered into 
between third parties and any Credit Support Providers\footnote{A Credit Support Provider is any party designated within the Master Agreement as providing some form of 
credit support to one of the parties (e.g. a guarantee)} 
or Specified Entities\footnote{Parties may extend the application of certain Events of Default (Default under Specified Transaction, Cross-Default, Bankruptcy and Credit Event upon Merger) and a Termination Event (Credit Event upon Merger) to certain other ``Specified Entities.''}
of the respective parties. 
A scenario may therefore arise
where this Event might occur due to circumstances existing with 
respect to arrangements entered into between two parties, neither of whom are parties to the Master 
Agreement. Without continuing access to information or data around these arrangements, making a 
determination around whether the relevant circumstances have arisen may prove very challenging to 
automate.

\subsubsection{Level 4: Exterior Level.}
Much of the complexity at this level arises due to the large number of external events that may arise and 
the difficulty of determining those external events that may be relevant in determining when an Event has 
occurred under the Master Agreement. 

Some Master Agreement Events at this level (such as an Illegality and Force Majeure) 
are necessarily broad in scope so as to encompass the existence of a wide range of 
circumstances that may, for example, make it illegal or impossible for parties to fulfill their obligations.

Ongoing observation and interpretation of information relating to each of the legal and regulatory 
frameworks applicable to all parties is likely to prove both challenging and inefficient to automate.  
Furthermore, it is also likely to be challenging to automate the interpretation of the extent to which such an 
external event might at any point operate such as to 
apply (or not apply) a prohibition or restriction on one or more parties.  It may be particularly difficult where the interpretation
might lead to the restriction of an activity  that is necessary for the parties to continue to perform all obligations 
under the Master Agreement.

Despite the inherent difficulties, we believe it may be possible to automate some aspects of the monitoring of 
events at this level, perhaps with the
smart contract code monitoring some readily-available external information and providing alerts that will then
be followed by human interpretation.  In other situations, human observation of an external event may require
the ability to pause or stop the smart contract code (e.g. in the case of an Illegality).

\section{Automating payments and deliveries}

In the context of payments and deliveries, it is important to look at events occurring in all four framework levels.\footnote{
Here we are discussing payments and deliveries
obligations that arise through the 
normal life of transactions entered into under the Master Agreement.  
We do not touch upon the extent to which the Master Agreement operates to modify 
payment obligations following the effective designation of an Early Termination Date.}

When parties enter into derivatives transactions, they will typically agree the type of trade they are entering into and the relevant economic terms, including how and when 
payment or delivery obligations might arise. They may also agree upon any relevant disruption or adjustment events that are intended to address 
issues that might arise 
throughout the lifecycle of the Transaction and which may prevent the transaction from operating in the manner intended.  For example, if parties enter into an interest rate 
swap and the relevant rate ceases to be published, the parties will look to any disruption or fallback provisions specified in the contract to ascertain how this disruption should 
be addressed.

These economic terms relating to an individual transaction are typically set out in a Confirmation. 
As previously explained, we refer to this as the ``Transaction Level''.
There are a number of other legal obligations that exist at the broader Relationship Level (and at the Third Party and Exterior levels) that may impact upon (or create new) payment and/or delivery obligations. 

Legal obligations that exist at the Relationship Level are typically found in the Master Agreement.
For example, the netting provisions of the Master Agreement\footnote{Section 2(c).} may have the effect of determining that a 
single net amount is payable on any given day in respect of one 
(or more\footnote{If Multiple Transaction Payment Netting is applicable.}) transaction(s) rather than multiple gross payments. The netting provisions may therefore affect the quantum of payment(s) that may be due between the parties on any given day. 

The actual legal obligation to make a payment may be impacted by other contractual provisions that exist at the Relationship Level. For example, the occurrence of an Event of Default 
under the Master Agreement may, in some scenarios, allow parties to suspend their obligation to make payments when due.

This interaction between the Transaction Level and the Relationship Level creates multiple levels of potential payment obligations between the parties to a transaction, 
all of which must be considered in the context of automation.

Here we will explore some of these provisions and determine how and in what circumstances they might impact upon payment obligations arising from the agreed economic terms 
at the Transaction Level. 

\subsection{Payment Obligation: Section 2(a)(i)}
The parties' respective obligation to make payments or deliveries to each other exists at the Relationship Level (i.e. in the Master Agreement):
\begin{quote}
{\em ``Each party will make each payment or delivery specified in each Confirmation to be made by it, subject to the other provisions under this Agreement.''}
\end{quote}

\noindent
The parties are required to make each payment or delivery specified in each transaction confirmation. It is therefore necessary to look to the economic terms of each individual transaction to determine when a payment or delivery is due and the relevant quantum of any payment of delivery. However, the obligation to make payments is subject to the other provisions existing 
within the Master Agreement.
Section 2(a)(i) 
therefore neatly illustrates the relationship that exists between payment obligations arising at the Transaction Level and at the Relationship Level, 
and highlights the importance of the Relationship Level to automation and to the development of the smart contract code.

\subsection{Single Agreement: Section 1(c) and Payment Netting}
By entering into the Master Agreement, the parties agree that:
\begin{quote}
{\em ``All Transactions are entered into in reliance on the fact that this Master Agreement and all Confirmations form a single agreement between the parties...''}
\end{quote}

\noindent
Therefore, transactions entered into under the Master Agreement do not create separate and distinct legal contracts between the parties. 
Instead, they are incorporated by reference into a single agreement under the Master Agreement architecture. 

One of the benefits of the single agreement architecture is the ability to net payment obligations arising under multiple Transactions at the individual Transaction level in order to determine a net sum which is then payable.

The payment netting provisions operate by determining the aggregate amounts due and payable by each party to the other  and determining that those payment obligations are replaced by a single, net payment obligation totaling an amount equal to the excess of the larger aggregate amount over the smaller aggregate amount.  This assists risk management and liquidity management, and is an important consideration for automation.

In order to automate payment netting, it will be necessary to validate the smart contract code, taking into account the relevant terms of the Master Agreement, any relevant negotiated modifications in the Schedule, and possibly additional provisions that might be contained in the Confirmation for each Transaction.  This may lead to the creation of multiple payment ``netting groups'' (groups of Transactions whose payments may be netted with each other); as new Transactions are created and existing Transactions are terminated, the number of Transactions within each netting group will vary, and
so the smart contract code must be able to manage these groups across extended periods of time.

\subsection{Condition Precedent: Section 2(a)(iii)}
The Master Agreement provides that the respective payment obligations of the parties are subject to:
\begin{quote}
{\em ``the condition precedent that no Event of Default or Potential Event of Default with respect to the other party has occurred and is continuing.''}
\end{quote}

\noindent
Therefore, the {\em performance} of a payment obligation depends on the condition precedent in Section 2(a)(iii) of the Master Agreement being satisfied, and it is important that the smart contract code is capable of monitoring (Potential) Events of Default (either directly, or via human input) and suspend outgoing payments where appropriate.

It is important to emphasize that Section 2(a)(iii) does not have the effect of expunging a party's payment obligation should an Event of Default (or Potential Event of Default\footnote{Defined in Section 14 
as ``any event which, with the giving of notice or the lapse of time or both, would constitute an Event of Default.''}) have occurred with respect to the other party. Rather, the payment obligation is {\em suspended} 
while the Event of Default  is continuing. This is the case even where the Event of Default\footnote{This is, of course, also the case for Potential Events of Default. However, given that Potential Events of Default are, by their very nature, time limited, discussion as to the potential duration of any suspension under Section 2(a)(iii) is perhaps less relevant.} is incapable of being cured, meaning the period of suspension may continue indefinitely.\footnote{Lomas v JFB Firth Rixson Inc [2012] EWCA Civ 419 (3 April 2012).}
In the Lomas v Firth Rixson case\footnote{Edward Murray. "Lomas v Firth Rixson: 'As you were!'". Capital Markets Law Journal. 8: 395. Retrieved 3 September 2015.} the Court of Appeal confirmed that a payment obligation of the Non-defaulting Party suspended as a result of the failure of the condition precedent continues indefinitely as a contingent payment obligation and may only be brought to an end by the curing of the relevant Event of Default or Potential Event of Default or the subsequent designation of an Early Termination Date.

While this position appears settled from an English law perspective, it is important to note that other jurisdictions may interpret the operation of Section 2(a)(iii) differently,\footnote{See, for example, the Metavante case (re Lehman Brothers Holdings, Inc., Case No. 08-13555 et seq. (JMP) (jointly administered) (2010).) where the Bankruptcy Court for the Southern District of New York held that the non-defaulting party could not rely on the safe harbour provisions in the Bankruptcy Code to excuse its failure to perform its obligations under the Swap Agreement. Therefore, it is uncertain as to whether parties can rely on Section 2(a)(iii) to suspend performance of payment obligations where a bankruptcy Event of Default has occurred with respect to the other party under Section 5(a)(vii) of an Master Agreement that is governed by New York law.} and therefore smart contract code may need to take into account different jurisdictional differences.

\subsection{Interest and Compensation: Section 9(h)(i)}

Where a payment is not made prior to the occurrence or effective designation of an Early Termination Date, whether because it has been deferred due to the operation of Section 2(a)(iii) or because a party has defaulted in the performance of any payment obligation, Section 9(h)(i) provides that interest or compensation may be due in respect of these defaulted (or deferred) payments or deliveries.

%
%

The calculation of interest or compensation could be calculated and effected automatically by the smart contract code.  However, although a party may have the right to apply interest or request compensation it may choose not to exercise that right.  For example, small amounts of interest might be waived for a favoured client.  This is a sensitive issue for automation and requires the capability for smart contract code to be able to communicate with one or more parties to obtain authorisation to proceed with the calculation and operation of these additional payments (and if it is not possible for the smart contract code to make the payments --- perhaps due to lack of funds in a designated account --- then the smart contract code may need the capability to raise the issue again with one or more parties to determine how to proceed).

\subsection{Multibranch: Section 10}
The Master Agreement provides that parties may enter into Transactions through different branches or offices within the same legal entity.

If parties wish to do so, it is possible to designate a party as a ``Multibranch Party'' in the Schedule to the Master Agreement. That party may then enter into transactions through, book transactions in, 
and make and receive payments and deliveries with respect to a Transaction through, any of the offices it lists in the Schedule to the Master Agreement.
If a party enters into a Transaction through an office other than its head office, the parties may agree in the Schedule the other party will have recourse to such party's 
head office as if the Transaction had been entered into through its head office.\footnote{Section 10(a).}

Thus, smart contract code must for example be able to recognise that payments may arrive from multiple sources and that any of these payments could be effective in discharging a contractual obligation to make payment.

\subsection{Tax: Section 2(d)(i)}
The Master Agreement requires all payments to be made:
\begin{quote}
{\em ``without any deduction or withholding for or on account of any Tax unless such deduction of withholding is required by any applicable law, as modified by the practice of any relevant 
governmental revenue authority, then in effect.''}\footnote{Section 2(d)(i).}
\end{quote}

\noindent
This provides that payment can be modified to the extent that there is a withholding or deduction that is required by applicable law. The Master Agreement then sets out the process for determining how, when and, most importantly, by whom, any relevant tax might be deducted or withheld.  These matters could be determined at the Transaction Level (e.g. specifed in the Confirmation) and tax could thereafter be calculated and deducted automatically by the smart contract code.

If, for example, any withholding tax does apply to a payment under the Master Agreement, the financial burden of that withholding tax is allocated through a gross-up provision within the Master Agreement.\footnote{Section 2(d)(i)(4).} Generally speaking, the tax burden will fall on the payer when the tax is any tax other than a tax imposed by reason of a present or former connection between the payee and any relevant taxing authority.\footnote{Defined in Section 14 as an ``Indemnifiable Tax''.} Otherwise, the tax burden will fall on the payee.

This paper does not intend to analyse in an exhaustive fashion all of the various instances in which a tax requirement levied by a relevant tax authority on either (or both) of the parties may result in a payment obligation being modified.
However, in the context of automation, it is important to understand that the Master Agreement does require parties to perform an assessment of their respective tax obligation and to ensure that any requirements to deduct or withhold for tax are duly communicated to the other party and taken account of when making payments, and this requirement will also apply to the development of smart contract code for the automation of each Transaction. It will also be important to identify the nature and purpose of any potential gross-up payment and the smart contract code could then automate the creation of  appropriate records to account for tax and to facilitate audit processes.

 \newpage
\subsection{Managing events}
\label{sec:managingevents}

We conclude this section by considering what facilities must be developed within smart contract code to support effective automation, and especially automation of the processing of events.

Effective processing of events will require the following steps:
\begin{enumerate}
\item
Observation
\item
Determination
\item
Action
\end{enumerate}
 
 \noindent
Each of these is discussed below.  Effective processing will also require appropriate interaction with humans: either to receive information about events from humans, or to ask humans for authorisation to take certain discretionary steps, or to receive directions from humans with regards to the current operation of the smart contract (e.g. to pause or terminate the smart contract code, or to update the details relating to a party due to an assignment, and so on).

\subsubsection{Observation.}
The first step in processing events is the ability to observe: where the performance of a Smart Derivatives Contract is automated via smart contract code, all parties delegate some of this observation to the smart contract code, but we envisage that some observation will also remain with the parties who must then be able to communicate observed events (and other matters) to the smart contract code.

Observation breaks down into two aspects: what to observe, and how to observe.  These are linked: for example, some events may arise within the technology platform and are relatively straightforward for smart contract code to observe, whereas events arising externally may be more difficult to observe 
(for example, with a distributed ledger platform, an ``oracle'' must be established in advance to make the external observation and route it through to  multiple instantiations of the smart contract code (so that they all receive identical information).  

%

In some situations observation may be simplified by the Master Agreement's requirement that
upon the occurrence of a Termination Event (though not for an Event of Default),
the Affected Party must promptly notify the other party upon becoming aware of it and provide 
information to the other party which specifies the nature of the Termination Event, which derivatives 
transactions are Affected Transactions\footnote{An ``Affected Transaction'' is any derivatives transaction that is
considered to be impacted or affected by the occurrence of a particular Termination Event. It may be that all
derivatives transactions are affected by the particular Termination Event (See Section 14).} 
\footnote{Note that both parties are required to notify the other party upon becoming aware of the occurrence of
a Force Majeure Event.}
and any other information that the other party might reasonably 
require. In the context of Smart Derivatives Contracts, either the notifying party or a notified party can convey
this information to the smart contract code (but only if the smart contract code has the ability to receive such notifications).

However, a party may not always be aware of a Termination Event 
irrespective of 
whether that party is the Affected or non-Affected Party (e.g. an Illegality may technically occur without either 
party being immediately aware).  In this case the ability for the smart contract code to observe a wide variety of event at the 
Exterior Level would be helpful.

If the smart contract code were to observe an event that can 
clearly be interpreted as a Termination Event then the smart contract code must have the capability to notify all 
parties so that more complex analysis (e.g. of affected transactions) can be made.

\subsubsection{Determination.}
Once an event or circumstance has been observed, the smart contract code must be able to determine whether or not the criteria for 
triggering an Event (either an Event of Default or a Termination Event) might be fulfilled.  
This requires the smart contract code to obtain and monitor information and understand 
the implication of that information as it relates to the precise circumstances that may ultimately constitute or 
give rise to the occurrence of a particular Event.  Of course, for computer code ``understanding the implication'' of a set of observed
events means that the mechanism and thresholds for such determination must be analysed in advance and incorporated into the smart contract code.

Each Event of Default and Termination Event set out in the Master Agreement has different 
mechanics governing its exercise.
For example, certain Events of Default can only occur once a specified 
grace period has elapsed. The existence of a grace period allows a party to take steps to cure the underlying 
event or circumstance giving rise to a potential Event of Default or Termination Event prior to that Event of 
Default or Termination Event actually occurring.

Determining the duration of the grace period upon the occurrence of a particular Event may not always be 
straightforward, and so the smart contract code must be subtle in this regard. 
Different durations are used for different types of Event.\footnote{There may also be differences across different 
forms of Master Agreement between the duration of grace periods in respect of the same event. 
Parties may also bilaterally agree to amend their contracts so that different durations apply.}
Some grace periods use calendar 
days for determining their duration, whereas others are determined by reference to days on which commercial banks 
are open in a relevant jurisdiction in which the parties are located (defined in the Master Agreement as 
``Local Business Days'').

While in most cases objective criteria are used in determining whether or not a relevant Event has occurred, 
the determination of some Events may include subjective elements. 
For example, a Credit Event upon Merger\footnote{Section 5(b)(v).}
may occur where a 
party is subject to a merger, acquisition or capital restructuring and the creditworthiness of the resulting entity 
becomes ``materially weaker'' as result. The Master Agreement does not provide any definition, 
explanation or guidance about what is meant by ``materially weaker''.  Where seeking to trigger this 
Termination Event, a party must therefore rely upon their own subjective interpretation of what is meant by 
``materially weaker'' in the context of their own contractual relationship and convey this information to
the smart contract code.

\subsubsection{Action}
When the circumstances giving rise to a potential Event have occurred and are continuing, the parties may be 
entitled to exercise certain contractual rights under the Master Agreement. 

A party may 
wish to terminate their contractual 
relationship with their counterparty. Alternatively, they may decide that the Event is relatively immaterial or 
inconsequential and that they do not wish to take any action.  As discussed above, payment and delivery 
obligations are contingent upon the condition precedent that no Event of Default or Termination Event with 
respect to the other party has occurred and is continuing.\footnote{Section 2(a)(iii)}
So a party may
decide to suspend payment obligations until the relevant Event is cured or is no 
longer continuing.  Therefore, 
there will often be uncertainty as to what the exact consequences of an Event will be due to the levels of 
human intervention and discretion required.\footnote{In some scenarios, Automatic Early Termination may apply 
meaning that upon the occurrence of certain types of 
insolvency related events, the Master Agreement will terminate automatically.}

It is unlikely that all counterparties will have identical appetites for risk, and therefore unlikely that they will all wish the
consequences of an Event to be managed in the same way.  Thus, it would seem that the default action for smart contract code
to take once an Event has been determined should be to inform the relevant parties and await further authorisation (though
for greater efficiency this
should be structured, so that for each Event a human can authorise one of a selection of pre-programmed further actions).

Looking ahead, it might be possible for smart contract code to have pre-programmed actions that are different for each party.
Since the smart contract code must be authorised by all parties, these pre-programmed responses will of course be known to all 
parties in advance, so some care will be required to ensure that these known responses cannot be exploited to the advantage of
defaulting party.  More subtle schemes might be imagined --- for example, the smart contract code could be instructed to observe 
a rising level of smaller events and thereby infer a rising level of risk, so that as the
risk grows the automated response to each subsequent Event becomes less lenient (or perhaps triggers an alert to the party at 
growing risk).


\section{Discussion}
\label{sec:discussion}


Here we discuss a number of issues that arise during consideration of
using Smart Dervivatives Contracts to automate payments and deliveries, and how these are affected
by the Master Agreement.  We cover two general issues of drafting precision and validation,
a more detailed issue of functionality relating to human intervention, 
and the strategic issue
of how much to automate.

\subsection{General issues}
\label{sec:generalissues}
Here we make some general comments to provide further clarification relating to two issues (drafting precision and validation)
that have been raised in other publications (specifically
\cite{ISDA2019}
and
\cite{ISDAKingWoodMallesons2018}).

\subsubsection{Drafting precision.}

There is some discussion in \cite{ISDAKingWoodMallesons2018} of the difference in ``drafting precision''
between the human language used by lawyers and the formal programming languages used to create
computer code.  The paper states:

\begin{quote}{\em
``a human user has more flexibility to work out the right method to follow to understand legal drafting 
than a machine has to understand the right method in programming languages.''
}\end{quote}

\noindent
This might be taken to imply that computers and computer programming are necessarily deterministic, or are unable to manage ambiguity, 
neither of which are true.  Although it is true that modern computer
hardware has been designed to be deterministic, and that programming is often deterministic, this is for
 human convenience --- because through that determinism we achieve control over the hardware.
It is however straightforward to build a computer that is non-deterministic,\footnote{Perhaps to implement a non-deterministic finite automaton:  http://mlwiki.org/index.php/Non-Deterministic\_Finite\_Automata} or 
to create a software layer that adds a level of non-determinism and to program in a non-deterministic way.\footnote{Non-deterministic programming is a well-known
sub-discipline of computer science. See https://en.wikipedia.org/wiki/Nondeterministic\_programming.}  Furthermore, it would be
reasonably straightforward to create a compiler\footnote{A ``compiler'' translates from human-readable programming language 
 source code to computer-readable machine code.}  that would
 ``guess'' at the meaning of any source code that was not recognised, or to create computer systems
 that explore different versions of a program until the best fit is found to a desired outcome (e.g. Genetic Programming \cite{koza1994genetic,yan2006behavioural,yan2008learning}), and it is straightforward for computer programs to utilise many-valued logics: for example, a three-valued logic with truth values {\em True}, {\em False} and {\em Unknown}, or a logic with potentially infinitely many values, such as ``fuzzy logic'' and ``probability logic''.\footnote{See https://en.wikipedia.org/wiki/Many-valued\_logic}


Thus computers need not be deterministic, and need not be constrained to simple logic  --- but to what extent is this required or desirable in order to
automate derivatives contracts?  A similar question applies to drafting legal text using English:
how much of the text should be absolutely clear and unambiguous in advance, and how much can be left to later interpretation?
It is interesting to observe that legal contracts have a {\em mixture} of those aspects that should be very
closely and carefully specified in advance, with no ambiguity, and those aspects that are expressed with a degree
of ambiguity either because (i) they are unlikely to occur, and are not worth the cost required to express them precisely, or
(ii) the partes are unable to agree precise terms but are happy to agree ambiguous terms.  Furthermore, discretion can be a key aspect of responding to
events (or potential events) of default, and although we could envisage imbuing the smart contract code with a degree of artificial intelligence to
make discretionary choices on the behalf of the parties, those parties might not wish to adopt this approach for high-value, 
complex financial products such as derivatives.


\subsubsection{Validation.}
Both
 \cite{ISDA2019} 
 and
 \cite{ISDAKingWoodMallesons2018}
 emphasise the importance of validation, reinforcing and expanding on the views previously 
 expressed in \cite{SCT-R3-3,ClackJDB,SCT2016,SCT2016a}.  For example:
 
\begin{quote}{\em
``It is important to ensure lawyers are able to validate that the legal effect of any coded or
automated provision is certain, and that the legal effect of the code aligns with the intended
legal effect of the contract.

This is particularly important in the derivatives market, given that derivatives contracts are
often used in connection with each other. For example, one derivatives contract may be used
to hedge the financial exposure created by another. An inability to validate the legal effect of a
smart derivatives contract may therefore introduce increased risks for the parties and the wider
derivatives market.'' \cite{ISDA2019} 
}\end{quote}

 \noindent
There are however many aspects to validation.  First, there is a particular kind of validation that 
must take place before any code is written, and this is to validate how smart contract code is  
viewed in law.  As an essential pre-requisite for automating certain kinds of action using smart 
contract code, we must understand in advance how the courts will interpret those actions, so 
that their legal effect is certain;  examples include the transfer of ownership of assets
on a distributed ledger, and whether pure crypto assets are property that can be owned.
For the ``internal model'' (see Section~\ref{sec:smartcontractcode}), there is a further essential 
pre-requisite which is to determine whether the courts would accept the legal force of
contractual provisions expressed using a formal representation (such as code).
These are examples
of advance validations of the technology and its intended operation, and must be tested in multiple jurisdictions.

The second kind of validation is that which we have mentioned previously in Section~\ref{sec:smartcontractcode},
which is to validate the smart contract code that has been developed to automate the performance of
a given Transaction.  Without the standardisation provided by the ISDA document architecture, this might
be a lengthy and expensive task that would have to be repeated for each Transaction; happily, the standardised 
and structured approach provided by the ISDA legal documentation means that much of the chore of
verification and validation of code can be undertaken early in the software lifecycle and therefore much of
this effort can be shared between agreements and between Transactions (this is the purpose of
Smart Contract Templates, as explained in Section~\ref{sec:smartcontractcode}).

This second kind of validation has the overall aim of validating whether the operation of the smart 
contract code (under all possible scenarios) is faithful to the intentions of the parties and to the legal 
effect of the contract and its governing law.  It must also ensure that the mechanisms for stopping or
pausing and modifying the smart contract code are sufficient and effective, and that there has been
correct identification and effective implementation of all situations where the code should pause and 
contact one or more parties for further input and authorisation regarding how to proceed.  Specifically,
it will be important to validate whether (i) the operation of, and (ii) the legal effect of, the smart contract 
code are {\em certain} where they need to be certain, and whether the way in which particular actions 
have been implemented in the smart contract code has the intended legal effect under all possible 
conditions.\footnote{\cite{ISDAKingWoodMallesons2018} provides an example where 
a particular technical operation of close-out netting might not be legally effective.}

In \cite{ISDAKingWoodMallesons2018} there is mention of a third, lower-level, kind of validation, which
is to check the operation of low-level aggregations of functions (in \cite{ISDAKingWoodMallesons2018} these are called
 ``templates'', which should not be confused with Smart Contract Templates).  These low-level procedures are intended to
be shared by many different derivatives products and many different Transactions under many different
agreements; therefore, it is not possible to validate whether these procedures implement the
intentions of the parties (who are unknown) nor whether they correctly implement a product (which is unknown).
What is required instead is the design of a formally-specified system for changing the internal state of a contract,
where these low-level procedures implement the changes in state and can be validated in terms of formal
notions of correctness --- and the system can be formally validated in terms of correctness, completeness
and consistency.

 
\subsection{Human Intervention}
As discussed previously in Section~\ref{sec:managingevents}, it is essential for smart contract code to support human intervention 
in terms of:
\begin{enumerate}
\item
permitting a party to give external direction such as to pause or stop the code (e.g. where an action that would be taken by the code has become illegal, or when the parties
have agreed to terminate the contract prematurely);
\item
receiving external data from a party where there is no suitable digital source for the required information; and
\item
pausing its own actions to request further guidance/authorisation from a party
(e.g. where the determination of whether an Event has occurred requires a party to make a subjective judgement or assessment).
\end{enumerate}

\noindent
If the smart contract code were to run on a distributed ledger, there would normally be multiple running copies of that code and any support for
external direction of the code must be done in a way that is simultaneously effective for all running copies of the smart contract code.
Similarly, if one copy of the smart contract code needs to pause to request human input it should be the case that {\em all} copies of
the code will request that input --- in this case, only one input should be required and that input should be simultaneously transmitted to all copies
of the code.  A recognised way to achieve synchronisation between incoming information from the external world and multiple running copies of smart contract code is to use ``oracles'' --- a single oracle will fetch data from the external world and place (possibly time-stamped) values on the blockchain,
thereby making identical data available to all copies of the smart contract code.  A similar technique could be used for outgoing synchronisation: multiple running copies of the smart contract code could agree to place on the blockchain a datum that indicates a request for input from a specified party: all copies of the smart contract code would pause; an oracle would read the request and contact the appropriate party, then receive the human response and place it on the blockchain for all copies of the smart contract code to read and determine their next action.

In relation to the premature termination of smart contract code due to human intervention, 
we should consider whether the ``internal'' and ``external'' models are affected in the same 
way.  As stated by \cite{ISDAKingWoodMallesons2018}, the premature termination of the code does not by itself suspend the contract but merely its automatic performance, and this 
should be the same for both models.  
{\colo{red}But in the internal model, would the code that has been terminated be viewed as part of the contract such that stopping the code might be deemed to be a cancellation of that part of the contract, or might give one or all of the parties a right to cancel the contract?  In the absence of any specific provision that contemplates this situation, we suggest that neither of these views should necessarily be true.  Only the automated performance has been prematurely terminated, and the formal representations of the internal model may continue to represent the agreement of the parties and to express the terms of that agreement (from a computer-science perspective, what has been terminated is the process, not the program). To the extent that the parties are able to continue meeting their respective obligations as agreed and expressed under the contract (whether through the resumption of the code or through some other means), we feel that the premature termination of the code should not necessarily give rise either to automatic termination or to a right to terminate the contract. }

\subsection{How much to automate}
\label{sec:howmuchtoautomate}
It has been repeatedly observed that in automating the performance of Smart Derivatives Contracts it is unlikely that the entirety of the
legal contract will be converted into smart contract code \cite{SCT-R3-3,ClackJDB,SCT2016,harley2017,ISDAKingWoodMallesons2018,ISDALinklaters2017} and so it is important to choose which
provisions shall be automated.  As observed by \cite{ISDAKingWoodMallesons2018}, it is important to consider both (i) what can be 
automated, and (ii) what should be automated.  This of course is not a static consideration: the former will increase for example as 
we gain a better understanding of contract semantics
and as technology improves, and the latter will vary for example according to jurisdiction and legal certainty and the appetites of the parties.

%

\subsubsection{What can be automated?}

In 2015--2016 early research into Smart Derivatives Contracts and Smart Contract Templates \cite{SCT2016} 
identified the {\em operational} and {\em non-operational} aspects of a contract.  The adjective ``operational'' 
referred to those aspects of the contract that are concerned with actions --- with the doing of things (such as 
exercising rights and discharging obligations) during performance of the contract.
The noun ``aspect'' was  chosen carefully to avoid being linked to sentences, paragraphs, clauses or provisions 
(since it was already clear that semantic concepts do not necessarily cleave to these simple syntactic structures 
in a one-to-one correspondence).   They were roughly characterised as the parts {\em ``that we wish to automate''} 
and the parts that {\em ``we do not wish to (or cannot) automate''}.  It was envisaged that: {\em ``The semantics 
of the non-operational aspects of even quite straightforward contracts can be very large and complex, yet by 
contrast the semantics of the operational aspects might be simple and easily encoded for automation''} \cite{SCT2016}.

Unfortunately, the matter has proven to be not quite that straightforward:

\begin{enumerate}
\item
It is not necessarily the case that an ``operational'' aspect is easier to automate than a ``non-operational'' aspect.
For example:
\begin{itemize}
\item
The following operational phrase (taken from an example Confirmation document) requires quite complex
code: {\em `` ``Strike'' shall mean the single Strike Price on both Underlying Option 1 and Underlying Option 2, expressed as the number of Yen per one United States Dollar, such that the net long/short of the delta of Underlying Option 1, and the delta of Underlying Option 2, using the Black-Scholes model, is neutral, i.e. the Delta Neutral Strike.  The Strike Price will be determined with consideration of the prevailing market rates on the Effective Date with respect to spot, forward points, interest rates and using an implied volatility of [130]\%.''}\footnote{Although written in a passive way, the final sentence is operational: it prescribes the exact volatility value that must be used during calculation of the Strike Price, and prescribes that three parameters (the values of spot rate, forward points, and interest rates) must be used in calculation of the Strike Price.  The actual expression to calculate the Strike Price will be given elsewhere in the economic terms or product definitions, and this provision prescribes operational constraints on that calculation.}
\item
By contrast, Section 3(a)(i) of the Master Agreement is a non-operational representation that each party ``is duly organised and validly existing under the laws of the jurisdiction of its organisation or incorporation'', which might (depending on jurisdiction and availability of online records) quite simply be checked automatically whenever a new Transaction is entered into.
\end{itemize}
\item
It is not possible to identify whether a part of a contract is ``operational'' or ``non-operational''
simply by inspecting the text to determine whether it uses conditional logic\footnote{This refers to a common structure in coding --- the
conditional statement or expression ``if [test] then [result] else [alternative result]'' where the test is either true (leading to ``result'') or
false (leading to ``alternative result'').} (although this has been widely assumed).  
Some operational phrases do not use conditional logic,
and some phrases that use conditional logic are non-operational.  For example:
\begin{itemize}
\item
In the Confirmation for a Straddle (comprising two linked options) the following  statement may appear:
{\em ``Upon payment of the Premium, the Buyer may elect to exercise either Underlying 
Option.''}  This is clearly operational (it refers to the action of electing to exercise one
or other underlying option) yet it has no conditional expression --- there is no ``if $\ldots$ then $\ldots$ else $\ldots$''
construct.\footnote{Though it could be rephrased to use a conditional expression.}
\item
Section 5(b)(vi) of the Master Agreement states ``If any ``Additional Termination Event'' 
is specified in the Schedule or any
Confirmation as applying, the occurrence of such event $\ldots$'', and Section 9(b) of the Master Agreement states
``An amendment, modification or waiver in respect of this Agreement will only be effective if in writing $\ldots$''.
Both of these phrases use a conditional ``if'', yet neither is operational.
\end{itemize}

\end{enumerate}

\noindent
Furthermore from the outset it was assumed that most of the operational
aspects would be found in the Confirmation and product definitions and most of the 
non-operational aspects would be found in the Master Agreement.
However, further investigation revealed that the Master Agreement has a much 
larger operational effect than had been expected (as explored in this paper).  
An initial study of the semantics of the Master Agreement
revealed not only a large operational aspect, but also an unexpected entangling of deontic, temporal and
operational aspects \cite{ClackJDB}:
\begin{quote}{\em
It    would    greatly    encourage    the    programmers    of    smart    contract    code    if    it    were    possible    
to     identify    the purely     operational     parts     of     a     contract.          These     could     then     be     encoded    
without    requiring    the    programmer    to    consider    complex    deontic    or    temporal    aspects.        
However,   
[$\ldots$]
%
many     actions     have     embedded     temporal     aspects     (trivially)     and     may     have    
deontic     aspects     embedded     within     the     operational text.
}\end{quote}

\noindent
Further investigation 
identified three key
problems \cite{ClackVanca2018}:

\begin{quote}{
\begin{enumerate}
\item
The {\em separability problem} --- the temporal, deontic and operational logics are
closely intertwined and very difficult to separate 
\item
The {\em isomorphism problem} --- the structure of the semantic specification (or the code) may
be substantially different to the structure of the legal documentation, with several consequences:
\begin{itemize}
\item
Seemingly complex legal prose may sometimes be automated in a relatively simple way.
\item
Seemingly simple legal prose may sometimes require relatively complex automation.
\item
It may sometimes be difficult to validate the code because the structure of the code does not match the structure of the legal text.
\end{itemize}
\item
The {\em canonical form problem} --- {\colo{red}there may be many different ways to structure
the semantic specification for a given legal agreement; specifically it is not yet clear whether
it is possible to transform all such different structures into a unique standardised form (``canonical form''), making it difficult to
compare two specifications for equality.}
\end{enumerate}
}\end{quote}

%
%
%


\subsubsection{What should be automated?}
If automation were limited to the basic economic conditions outlined in the Confirmation and product definitions,
the accruing benefit would be modest in comparison to what could be achieved by also automating the provisions
of the Master Agreement.  A truly autonomous Smart Derivatives Contract should for example be 
capable of observing a range of events and detecting common (Potential) Events of Default. This
does not mean that the entirety of the Master Agreement must necessarily be automated, and it is important
to reason about which parts {\em should} be automated --- e.g. because they are easy to automate, or because
their automation although difficult would bring great benefit.

We have provided a framework to guide thinking about what should be automated, by considering the observability
of events at different levels.  Events at the Transaction Level are perhaps easiest to observe, those at the
Relationship Level are slightly less easy to observe but their automation could 
bring great benefit (e.g. payment netting).  Events at the Third Party and Exterior levels may be easy or difficult to 
observe --- some of them would require a ``vast number of complex and interdependent permutations'' to be considered
\cite{ISDA2019} whilst others might be relatively simple to observe.  The ease of observation at Third Party and
Exterior levels will vary according to jurisdiction and according to the maturity and interconectivity of the technology.

Although computers vastly surpass humans in their ability to manage huge amounts of data and complex inter-relationships,
it will require human effort in advance either to generate and classify that information or to validate data and inter-relationships 
that are discovered by automatic search.  Yet this level of effort may not be necesary if what is commonly known as the Pareto Principle \cite{juran1975non} is applicable
(i.e. that 80\% of the benefit will accrue from 20\% of the effort).

``The 20\%'' should not necessarily be interepreted as ``20\% of the legal provisions'' since in the
implementation of a single provision there may be parts that are easily automated, and parts that either are not easy to automate
or we do not wish to automate (perhaps because they require human interaction).  And the desirable extent of automation in
each case may vary.  Thus, ``the 20\%'' should be viewed as a general guideline to be applied flexibly --- to identify what
can be achieved autonomously, while maximising the ratio of benefit to effort.

Some guidelines have been suggested in \cite{ISDAKingWoodMallesons2018}, and we briefly
{\colo{red}paraphrase and} discuss these guidelines below:

\begin{quote}{
\begin{itemize}
\item
{\colo{red}{\em Standardisation: }
focus on automating common, standardised, aspects of derivatives 
contracts, so that the automation is widely applicable across a large number of different contracts.}

Standardisation is also important for another reason: so that there is time to thoroughly 
verify and validate the code in advance.  The efficiency of ``shareability with others'' 
happens via the use of Smart Contract Templates
({\colo{red}e.g.} code that automates some aspects of the Master Agreement can 
be shared and used when constructing code
for a large number of agreements each with a negotiated Schedule).

{\colo{red}Although a negotiated provision in a Schedule might be considered ``non-standard'',
in some cases it could be}
shared by a very large number of Transactions and its automation could potentially 
increase efficiency enormously (though it depends on how many Transactions derive from the relationship, 
how often the particular provision is activated, and the
alternative cost of manual activation).

\item
{\colo{red}{\em Complexity:}
avoid automating complex legal provisions, since these might be more difficult to establish, operate and maintain.}

We have observed that complex legal text can sometimes be captured with quite simple logic (and therefore
simple code).  The reverse is also true, that seemingly simple legal text may require quite
complex logic (and therefore complex code).  For example, Section 2(a)(iii) is a very straightforward
statement of general conditions precedent yet might require quite complex code, whereas Section 2(c) provides
a fairly complex explanation of payment netting whose semantics could be quite simple to automate.
This is the {\em isomorphism problem} mentioned above.
It is however generally true that simpler code is easier to 
verify, validate, operate, and maintain.

\item
{\colo{red}{\em Externalities:}
consider how external factors such as observable events or discretion (including by a third party) will be efficienty incorporated into the smart contract code.}

We concur with this guidance and have given some further discussion above relating to human
interaction, which we view as an essential part of smart contract code.

\item
{\colo{red}{\em Commonality [of ISDA CDM functions]:}
functions defined in the ISDA CDM should be common to many deriatives products.
}

{\colo{red}It is highly likely that many high-value} Smart Derivatives Contracts 
will be automated via smart contract code that  uses functions defined by the 
ISDA CDM {\colo{red}\cite{CDM2017}},  and it is  highly desirable that 
the ISDA CDM functions should be common across a wide range of {\colo{red}derivatives} products.
Commonality will not be the only criterion for ISDA CDM functions:
they should also, for example, be chosen carefully so that they are readily composable and so
that the set of available functions is concise.



\item
{\colo{red}{\em Validation of legal effect:}
only automate those aspects of a derivatives contract where a lawyer can confirm that
their legal effect will not be changed when automated.

This is very much the thrust of our approach: we view legal validation as} essential, not just of individual aspects but of the whole
smart contract code.
\end{itemize}
}\end{quote}

\noindent
The overall approach advised by \cite{ISDAKingWoodMallesons2018} is to pursue the following steps:
{\em
\begin{enumerate}
\item
Selecting parts of a derivatives contract for which automation would be effective and efficient;
\item
Changing the expression of the legal terms of those parts of the derivatives contract into a more 
formalized form; 
\item
Breaking the formalized expression into component parts for representation as functions;
\item
Combining the functions into templates for use with particular derivatives products; 
and
\item
Validating the templates as having the same legal effect as the legal terms of a derivatives contract.
\end{enumerate}
}

\noindent
We concur with all of the above.  The ``templates'' mentioned here are not the same as the Smart Contract
Templates introduced in Section~\ref{sec:smartcontracttemplates}, but rather are aggregations of
a relatively small number of ISDA CDM functions.
The smart contract code for a Smart Derivatives Contract
could utilise several of these ``templates'' as well as indivdual ISDA CDM functions, which would
manipulate aspects of the internal state of the smart contract code.

\section{Summary and Conclusion}

Smart Derivatives Contracts aim to automate high-value derivatives contracts, including automation of aspects of the Master Agreement 
 as well as automation of lifecycle events stated in the economic terms of the specific derivatives product.  
This vision raises many issues to be solved, such as  (i) how the smart contract code can be faithful to the legal agreement, and (ii) to what extent the provisions of the legal agreement can be automated.  This requires an inter-disciplinary approach that brings together computer scientists, lawyers and banking practitioners.

Much of the operational detail of payments and deliveries can be found in the transaction Confirmation and product definitions (i.e. the economic terms and payment mechanics of the particular derivatives product).  However it is not sufficient just to automate the operational aspects of the contract found in the economic terms; the broader contractual relationship must also be taken into account in order to capture the {\em ``complexity that can affect a party's ability to perform its obligations (or assert its rights)''} \cite{ISDA2019} and in order to support key operations such as netting that are expressed in the Master Agreement.  

%

Here we have investigated how the Master Agreement affects the automation of payments and deliveries: 
\begin{itemize}
\item
we have explained why it is not sufficient merely to automate the economic terms of a transaction, and how
the Master Agreement can directly affect both the timing and quantum of payments and deliveries (for example, with regards to
payments netting, and exercising rights to suspend payments);
\item
we have provided a framework for understanding how high-value derivatives contracts are structured at four different levels:
\begin{itemize}
\item
the Transaction Level;
\item
the Relationship Level;
\item
the Third Party Level; and 
\item
the Exterior Level.
\end{itemize}
\item
we have provided a framework of three steps for smart contract code to manage events relating to payments and deliveries: 
\begin{itemize}
\item
Observation;
\item
Detection; and 
\item
Action
\end{itemize}
\item
we have discussed the extent to which different kinds of events and processing are amenable to automation,
and explained that is is essential that smart contract code should support human intervention:
\begin{itemize}
\item
to permit human control such as to pause or stop the code;
\item
to receive external data where there is no suitable digital source;
\item
to request human guidance and/or authorisation.
\end{itemize}
\end{itemize}

\noindent
Developing smart contract code for deployment on a distributed ledger is a specialist function, requiring experienced technology practitioners and computer scientists (and in Section~\ref{sec:generalissues} we also briefly address the need for low-level ISDA CDM functions to be fully validated in advance 
so that they can be utilised when developing the smart contract code).  In the light of this specialist nature (and complexity), and in order to 
ensure that the code is faithful to the legal contract, we have explained that it is essential for development to be a co-operative effort involving 
(at least) the skills and experience of banking technology practitioners, legal professionals and computer scientists:
to address the legal context of the
operations that are being automated; to ensure that the code is compatible with the relevant commercial, regulatory and legal standards;
and to ensure that the code is consistent with the way that the parties wish to conduct their financial transactions.  

\newpage
\bibliographystyle{splncs04}
\bibliography{Clack-McGonagle-2018}
%
%
%
%
%
\end{document}